\documentclass[aps,twocolumn,prl,superscriptaddress]{revtex4-1}

\usepackage{amsmath}
\usepackage{amssymb}
\usepackage{color}
\usepackage{hyperref}
\usepackage[normalem]{ulem}

\usepackage{graphicx}

\DeclareMathAlphabet{\mathitbf}{OML}{cmm}{b}{it}

\newcommand{\rv}{\mathitbf r}

\newcommand{\calBold}[1]{\mbox{\boldmath${\cal #1}$}}
\newcommand{\mathBold}[1]{\mbox{\boldmath$#1$}}

\begin{document}
\title{Frustration-induced internal stresses are responsible\\ for quasilocalized modes in structural glasses}
\author{Edan Lerner}
\affiliation{Institute for Theoretical Physics, University of Amsterdam, Science Park 904, 1098 XH Amsterdam, The Netherlands}
\author{Eran Bouchbinder}
\affiliation{Chemical and Biological Physics Department, Weizmann Institute of Science, Rehovot 7610001, Israel}

\begin{abstract}
It has been recently shown [E.~Lerner, G.~D\"uring, and E.~Bouchbinder, Phys.~Rev.~Lett.~\textbf{117}, 035501 (2016)] that the non-phononic vibrational modes of structural glasses at low-frequencies~$\omega$ are quasi-localized and follow a universal density of states $D(\omega)\!\sim\!\omega^4$. Here we show that the gapless nature of the observed density of states depends on the existence of internal stresses which generically emerge in glasses due to frustration, thus elucidating a basic element underlying this universal behavior. Similarly to jammed particulate packings, low-frequency modes in structural glasses emerge from a balance between a local elasticity term and an internal stress term in the dynamical matrix, where the difference between them is orders of magnitude smaller than their typical magnitude. By artificially reducing the magnitude of internal stresses in a computer glass former in three dimensions, we show that a gap is formed in the density of states below which no vibrational modes exist, thus demonstrating the crucial importance of internal stresses. Finally, we show that while better annealing the glass upon cooling from the liquid state significantly reduces its internal stresses, the self-organizational processes during cooling render the gapless $D(\omega)\!\sim\!\omega^4$ density of state unaffected.
\end{abstract}

\maketitle


The statistical and structural properties of low-frequency vibrations in structural glasses have been the focus of many experimental~\cite{sokolov_boson_peak_scale,Monaco_prl2011,Buchenau_prb2014}, theoretical~\cite{soft_potential_model,Gurevich2007,eric_boson_peak_emt,silvio} and computational~\cite{barrat_3d,Schober_Laird_numerics_PRL,vincenzo_epl_2010,modes_prl, thermal_energies} research efforts for a few decades now. It is now established~\cite{modes_prl, thermal_energies, SciPost2016, protocol_letter, ikeda_pnas} and quite widely accepted that $(i)$ low-frequency glassy (non-phononic) vibrations in generic structural glasses in three dimensions follow a universal gapless density of states $D(\omega)\!\sim\!\omega^4$, where $\omega$ denotes the frequency, and $(ii)$ the modes that populate the $\omega^4$ regime are quasilocalized --- they feature a localized, disordered core that is either decorated by a power-law spatial decay away from the core \cite{modes_prl}, or accompanied by an elastic-wave-like background \cite{Schober_jop_2004,Schober_Ruocco_2004,ikeda_pnas}. The participation ratio (see precise definition below) of these modes scales as $N^{-1}$, where $N$ is the number of particles in the glass. Effective medium~\cite{eric_boson_peak_emt} and other mean-field theories~\cite{silvio} fail to capture this behavior of the vibrational density of states (vDOS); they predict instead $D(\omega)\!\sim\!\omega^2$, independently of the spatial dimension, and cannot (by construction) predict the quasilocalization of glassy modes.

The physical origin of the universal $\omega^4$ law in the vDOS of structural glasses is still a subject of current investigation. In~\cite{protocol_letter} it has been found that some extreme cooling conditions of computer glassy samples can give rise to vDOSs that grow as $\omega^\beta$ with $\beta\!<\!4$. Notwithstanding, in the same work it has also been shown that any physically realistic cooling rate (i.e.~in which heat is extracted at a rate that is much slower than the inverse vibrational time) results in a glass that exhibits the $\omega^4$ law, highlighting the role of the self-organizational processes that the glass undergoes as it explores lower energy states upon cooling.

In this brief report we focus on the role of frustration-induced internal stresses in forming the gapless $\omega^4$ vDOS observed in generic structural glasses. We show that a careful, artificial reduction of the degree of frustration-induced internal stresses in the glass leads to the opening of a gap in the vDOS, and the suppression of quasilocalized vibrational modes. In~\cite{ikeda_pnas} it was shown that a complete, artificial relaxation of internal stresses in harmonic soft spheres near unjamming leads to the disappearance of quasilocalized vibrational modes. Here we perform a quantitative study of the gradual suppression of quasilocalized vibrational modes as internal stresses are relieved. In the model studied, we find that reducing the internal stresses by merely $3\%$ leads to the elimination of the quasilocalized vibrational modes that populate the $\omega^4$ regime of the vDOS.

We employ a generic computer glass forming model: a 50:50 binary mixture of `large' and `small' particles in three dimensions (3D) that interact via an inverse power-law pairwise potential ($\sim\! 1/r^{10}$, with $r$ being the pairwise distance). A detailed description of the model can be found, e.g., in~\cite{protocol_letter}. Glassy samples are prepared by continuously cooling high-temperature liquids states at a rate of $\dot{T}\!=\!10^{-3}\epsilon/(k_BT\,\tau)$~\cite{protocol_letter}, where $\epsilon$ is a microscopic energy scale, $\tau$ is a microscopic time scale, and $k_B$ is Boltzmann's constant. In what follows all data are reported in terms of the relevant microscopic units, as defined in~\cite{protocol_letter}.

In our model system the potential energy is given by a sum over all pairs of particles, namely
\begin{equation}
U = \sum_\alpha\varphi_\alpha(r_\alpha)\,,
\end{equation}
where $\alpha$ labels pairs and $r_\alpha$ is the distance between the particles in each pair. In the harmonic approximation of the potential energy $U\!\simeq\!U_0\!+\!\frac{1}{2}{\mathBold x}\!\cdot\!{\calBold M}\!\cdot\!{\mathBold x}$, expressed in terms of the dynamical matrix ${\calBold M}\!\equiv\!\frac{\partial^2U}{\partial{\mathBold x}\partial{\mathBold x}}$ (where $\mathBold x$ is the $3N$-dimensional vector of particles' positions in 3D), any solid comprised of particles that interact via radially symmetric pairwise potentials can be thought of as a collection of masses connected by Hookean springs. The springs are characterized by stiffness (local elasticity spring constants) given by the actual interaction stiffness $\kappa_\alpha\!\equiv\!\varphi''_\alpha\!\ge\!0$, and rest-length determined by the internal forces (stresses) $f_\alpha\!\equiv\!-\varphi'_\alpha\!\ge\!0$. The emergence of frustration-induced internal forces (stresses) $f_\alpha$ is a generic property of glasses~\cite{shlomo}. The dynamical matrix is then expressed as~\cite{matthieu_thesis}
\begin{equation}
\label{eq:M}
{\calBold M} = \sum_\alpha\kappa_\alpha \frac{\partial r_\alpha}{\partial{\mathBold x}}\frac{\partial r_\alpha}{\partial{\mathBold x}} - \sum_\alpha f_\alpha\,r_\alpha^{-1}\frac{\partial^2r_\alpha}{\partial{\mathBold x}\partial{\mathBold x}}\,.
\end{equation}

We thus observe that the dynamical matrix ${\calBold M}$ is a difference between a contribution that involves a sum over the local elastic constants $\kappa_\alpha$ and a contribution that involves a sum over the internal forces (stresses) $f_\alpha$. Consequently, each vibrational frequency $\omega\!=\!\sqrt{{\mathBold \Psi}\cdot{\calBold M}\cdot{\mathBold \Psi}}$, where ${\mathBold \Psi}$ is the eigenvector corresponding to the eigenvalue $\omega^2$, can be expressed as $\omega\!=\!\sqrt{\omega_\kappa^2-\omega_f^2}$. We verified that low-frequency glassy modes, with $\omega\!\to\!0$ in the $\omega^4$ regime of the vDOS, emerge not because $\omega_\kappa^2$ and $\omega_f^2$ are each small, but rather because their difference is small, orders of magnitude smaller than their typical magnitude (data not shown). Similar observations have been made for jammed packings of soft spheres~\cite{silbert_pre_2016,ikeda_pnas}. This almost perfect balance between the local elasticity contribution $\omega_\kappa^2$ and the internal stresses contribution $\omega_f^2$ suggests that internal stresses play a crucial role in the emergence of the universal non-phononic $\omega^4$ law in the vDOS of structural glasses.

To directly test this idea, we artificially reduced the internal stresses in our glassy samples by defining a parameterized dynamical matrix ${\calBold M}(\delta)$ obtained from Eq.~\eqref{eq:M} by replacing $f_\alpha$ with $(1-\delta)f_\alpha$, where $0\!\le\!\delta\!\le\!1$. That is, these artificial glassy states correspond to the harmonic approximation of the original Hookean spring network in which the rest-length of each spring is altered so as to reduce the magnitude of the spring force by a factor $1-\delta$. The result of such an alteration would be to decrease the magnitude of internal stresses by exactly the same factor. The same procedure has been introduced and utilized numerically to reduce internal stresses in packings of soft harmonic spheres and disks near the unjamming point~\cite{eric_boson_peak_emt,breakdown}. Note that $\delta\!=\!0$ corresponds to the original glass and that $\delta\!>\!0$ alters both the eigenvalues $\omega^2$ and the eigenvectors ${\mathBold \Psi}$ of the artificial glass compared to the original glass.

\begin{figure}[!ht]
\centering
\includegraphics[width = 0.495\textwidth]{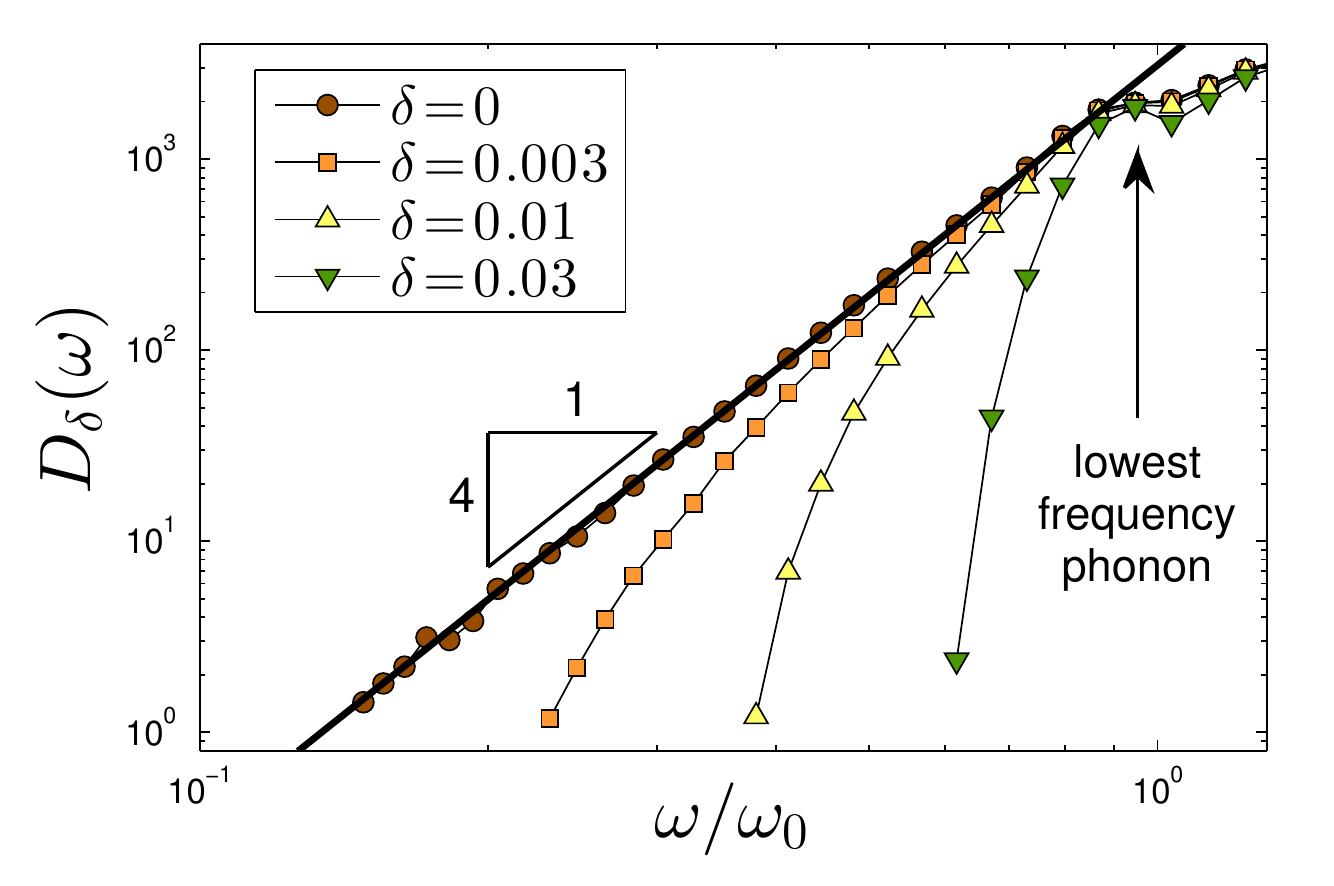}
\caption{\footnotesize Vibrational density of states $D_\delta(\omega)$ measured for ensembles of glassy samples characterized by the parameter $\delta$ that quantifies the arfiticial reduction of internal stresses, see text for its precise definition. We find that a gap opens up as the parameter $\delta$ is increased from zero. The scale $\omega_0\!=\!2$ was chosen for visualization purposes.}
\label{inst_fig}
\end{figure}
\vspace{0.3cm}

We calculated ${\calBold M}(\delta)$ in $50,000$ independently quenched glassy samples of $N\!=\!2000$ particles~\cite{modes_prl}, and analyzed the low-frequency regime of the corresponding vDOS $D_\delta(\omega)$. The results are presented in Fig.~\ref{inst_fig}. For the unaltered system we cleanly observe the $\omega^4$ scaling, i.e.~the spectrum is gapless. Once $\delta\!>\!0$ is set, we find a gap formed in the vDOS, and at $\delta$ as small as $0.03$, the $\omega^4$ regime of the vDOS is fully suppressed.

The fact that such a small reduction in the internal stresses is sufficient to suppress the $\omega^4$ regime of the vDOS can be understood in the framework of a simple perturbation theory to leading order in $\delta$. The perturbed (parameterized) dynamical matrix takes the form ${\calBold M}(\delta)\!=\!{\calBold M} + \Delta{\calBold M}(\delta)$, where the perturbation $\Delta{\calBold M}(\delta)$ satisfies ${\mathBold \Psi}\!\cdot\!\Delta{\calBold M}\!\cdot\!{\mathBold \Psi} \!=\!\omega_f^2\delta$. The eigenmodes are also perturbed in the form ${\mathBold \Psi} + \Delta{\mathBold \Psi}$, where $\Delta{\mathBold \Psi}\!\cdot\!{\mathBold \Psi}\!=\!0$ to leading order, due to the normalization condition $({\mathBold \Psi} + \Delta{\mathBold \Psi})\!\cdot\!({\mathBold \Psi} + \Delta{\mathBold \Psi})\!=\!1$. Finally, the perturbed eigenvalues take the form $\omega^2 + \Delta\omega^2$. 

\begin{figure*}[!ht]
\centering
\includegraphics[width = 0.95\textwidth]{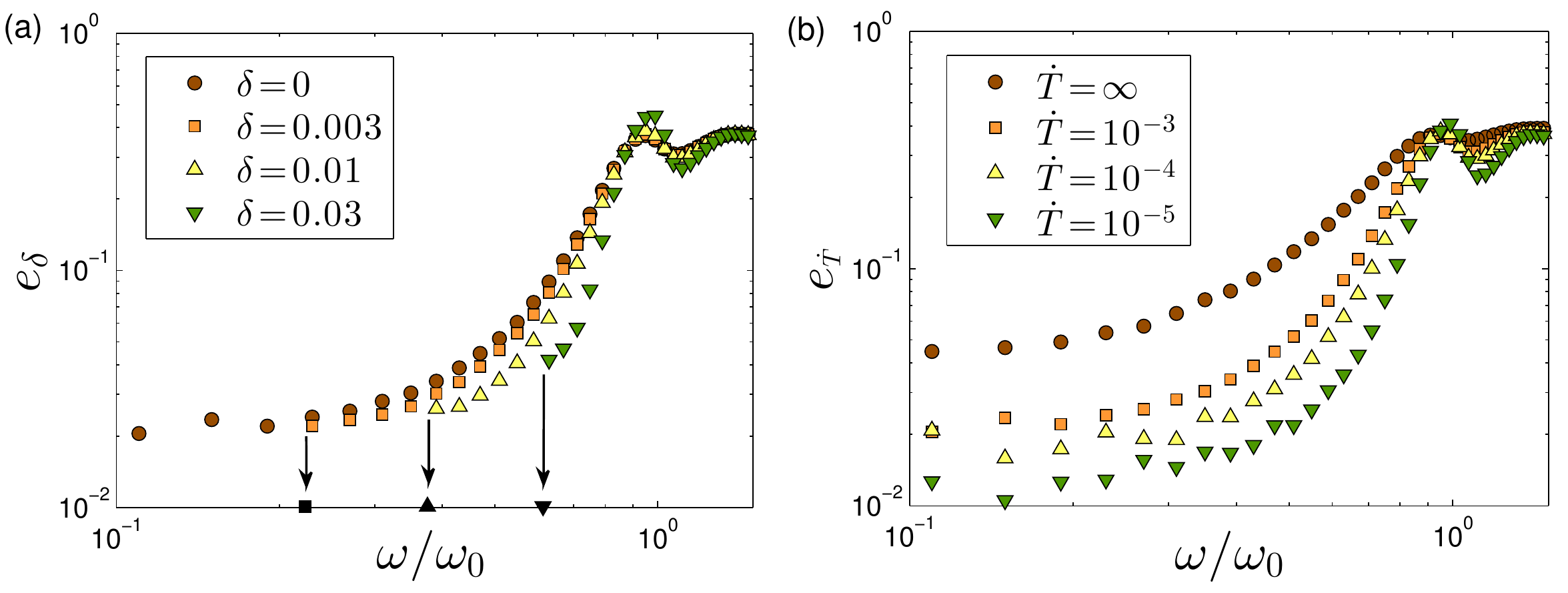}
\caption{\footnotesize (a) Mean participation ratios $e_\delta$ measured for vibrational modes calculated for systems in which the internal stresses are reduced by a factor $1-\delta$. The full symbols pointed at by the arrows mark the $\delta$-dependent truncation frequency, defined as the ensemble-lowest vibrational frequency for a given value of $\delta$. The scale $\omega_0\!=\!2$ was chosen for visualization purposes. (b) same as (a), but in ensembles of systems quenched from the high temperature liquid phase at a rate $\dot{T}$ as indicated by the legend.}
\label{inst_participation_fig}
\end{figure*}

The shift in the eigenvalues can be readily calculated from the following eigenvalue equation
\begin{eqnarray}
\omega^2 + \Delta\omega^2 & = & ({\mathBold \Psi} + \Delta{\mathBold \Psi})\cdot({\calBold M} + \Delta {\calBold M})\cdot({\mathBold \Psi} + \Delta{\mathBold \Psi}) \nonumber \\
& \simeq & {\mathBold \Psi}\cdot  {\calBold M}\cdot{\mathBold \Psi} + {\mathBold \Psi}\cdot \Delta {\calBold M}\cdot{\mathBold \Psi} = \omega^2 + \omega_f^2\delta  \nonumber \\
 &\Longrightarrow & \quad \Delta\omega^2 =  \omega_f^2\delta \ ,
\end{eqnarray}
where $\Delta{\mathBold \Psi}\cdot{\mathBold \Psi}\!=\!0$ has been used. Consequently, the analysis predicts that the eigenvalues $\omega^2$ shift upwards by $\omega_f^2\delta$ upon increasing $\delta$ from zero, and consequently a gap at vanishing frequencies that grows as $\omega_f\sqrt{\delta}$ should appear in the low-frequency tail of the vDOS. In our system we find $\omega_f\!\sim\!10$, and a lowest phonon frequency of order unity. We thus expect the $\omega^4$ tail of the vDOS to be fully suppressed when $\delta\!\sim\!10^{-2}$, as indeed seen in Fig.~\ref{inst_fig}.

We next consider the effect of artificially reducing internal stresses on the degree of localization of glassy modes. The latter is effectively quantified by considering the participation ratio of each mode ${\mathBold \Psi}$, defined as $e\!\equiv\!\big(N\sum_i ({\mathBold \Psi}_i\cdot{\mathBold \Psi}_i)^2\big)^{-1}$, where ${\mathBold \Psi}_i$ is the three-dimensional vector of the Cartesian components of ${\mathBold \Psi}$ pertaining to the $i^{\mbox{\tiny th}}$ particle. $e\!\sim\!{\cal O}(N^{-1})$ corresponds to quasilocalized glassy modes (in 3D) \cite{modes_prl, SciPost2016}, while $e\!\sim\!{\cal O}(1)$ corresponds to extended phonons (plane-waves). In fact, it is known that the presence of phonons with similar frequencies leads to their hybridization with glassy modes, i.e.~to a significant increase in their participation ratio~\cite{SciPost2016,phonon_widths}. In Fig.~\ref{inst_participation_fig}a we plot the mean participation ratio of vibrational modes $e_\delta$, binned over frequency, for systems parameterized by the same values of $\delta$ as shown in Fig.~\ref{inst_fig}. Interestingly, we find that artificially relieving the internal stresses leads to weaker hybridizations of glassy modes with phonons, as evident by the sharper increase of $e$ upon approaching the first phonon peak. We note that the truncation of the $e_\delta$ signals below a $\delta$-dependent frequency scale, marked by the arrows in Fig.~\ref{inst_participation_fig}a, is consistent with the truncation of the vDOS below the same frequency scale, as seen in Fig.~\ref{inst_fig}.

It is interesting to compare the effect of artificially reducing the internal stresses on the localization properties of soft glassy vibrational modes, to the effect observed by preparing glasses with slower cooling rates $\dot{T}$, which, in turn, also leads to a reduction of internal stresses. In Fig.~\ref{inst_participation_fig}b we show the results of the same analysis of the localization properties of low-frequency modes as shown in panel (a), applied this time to our thermally-annealed glasses. Details about the preparation of our ensembles of glassy samples can be found in \cite{protocol_letter}. We find a similar trend in these annealed ensembles to that observed in the artificially-stress-reduced samples: $e_{\mbox{\fontsize{4}{0}\selectfont $\dot{T}$}}$ increases more sharply towards the first phonon frequency in the better-annealed samples. A stark difference appears however in the frequency range in which the artificial reduction of internal stresses suppresses the existence of soft glassy modes: in the thermally annealed samples no apparent truncation is observed, i.e.~soft glassy modes appear to persist down to zero frequency, and their localization is stronger for better annealed samples.

\begin{figure}[!ht]
\centering
\includegraphics[width = 0.49\textwidth]{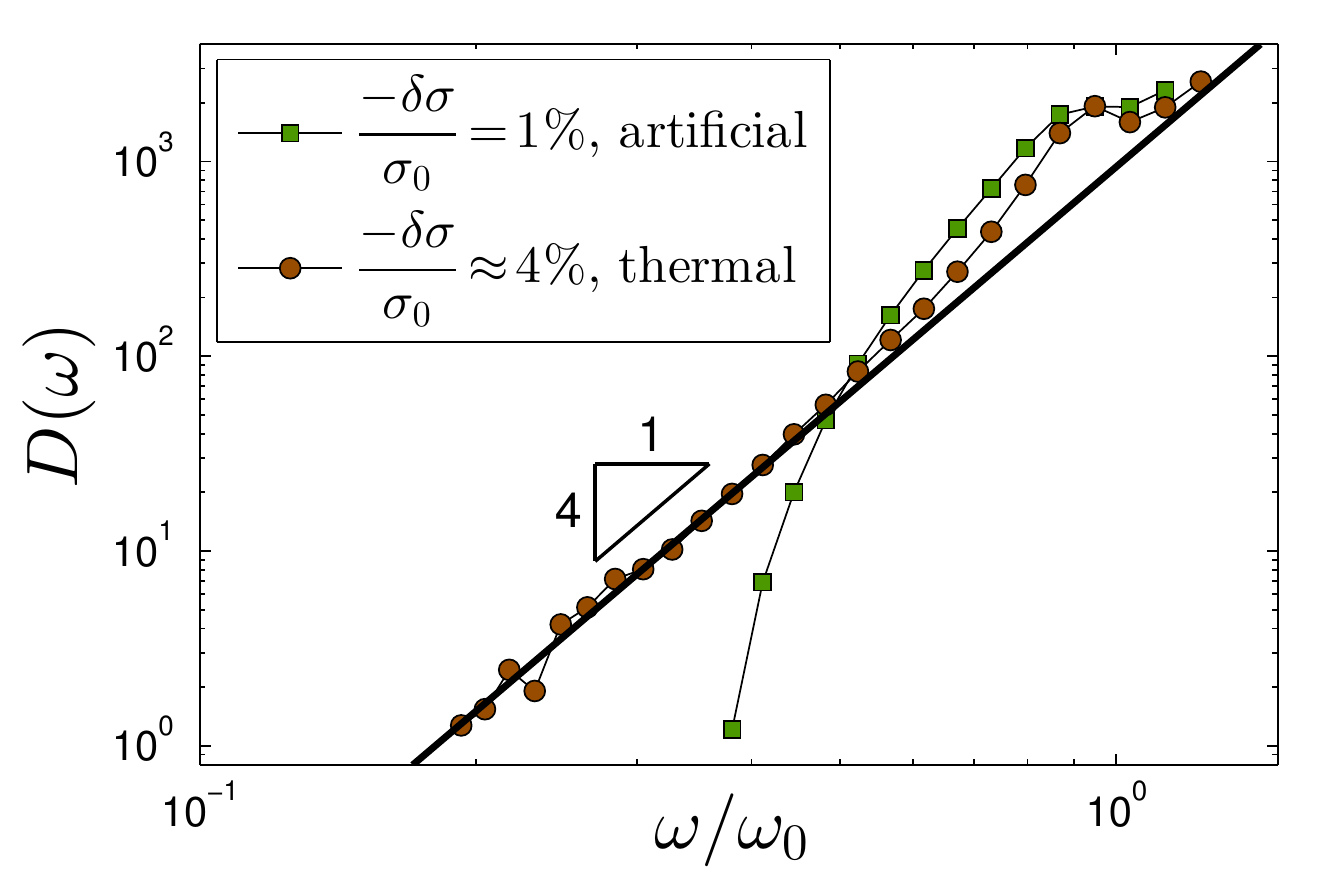}
\caption{\footnotesize Comparison of the density of vibrational modes of glassy samples in which internal stresses are reduced by a slower quench (thermally annealed), or by the $1-\delta$ procedure as described above (artificially annealed). The scale $\omega_0\!=\!2$ was chosen for visualization purposes.}
\label{compare}
\end{figure}


The aformentioned difference between the appearance of soft glassy modes in thermally-annealed \emph{vs.}~artificially stress-relieved samples can be directly gleaned from their respective vDOS. In Fig.~\ref{compare} we compare the vDOS of glassy samples in which internal stresses were reduced by a factor $1\!-\!\delta$ with $\delta\!=\!10^{-2}$ (green squares), to systems in which stresses are relieved by a slower quench, of $\dot{T}\!=\!10^{-5}\epsilon/(k_BT\,\tau)$ (brown circles). We quantify the magnitude $\sigma$ of internal stresses of each ensemble by measuring the sample-to-sample standard deviation of the $xy$ component of the stress tensor $-V^{-1}\!\sum_\alpha (f_\alpha/r_\alpha) \rv_\alpha\cdot\hat{x}\hat{y}\cdot\rv_\alpha$, with $V$ denoting the volume. The measured value of $\sigma$ of each ensemble is compared to that found in the ensemble shown in Fig.~\ref{inst_fig} for the reference glass (quenched at $\dot{T}\!=\!10^{-3}\epsilon/(k_BT\,\tau)$), $\delta\!=\!0$, denoted by $\sigma_0$ in the legend of Fig.~\ref{compare}. In the two compared ensembles, internal stresses are reduced by $\approx\! 4\%$ and $1\%$ for the better annealed glasses and the artificially stress-relieved glasses, respectively. Despite the substantially larger reduction of internal stresses in the annealed glasses, the spectrum of those glasses appears to maintain its gapless nature, whereas in the spectrum of the artificially stress-relieved glasses a gap is created. Consequently, the annealed glass possesses many more modes at the very lowest frequencies compared to the artificially stress-relieved glasses.

To summarize, we have shown in this brief report that artificially reducing the internal stresses in a model structural glass leads to the disappearance of low-frequency quasilocalized modes. This implies that the existence of low-frequency quasilocalized modes depends on the presence of such stresses, which are generic in glasses \cite{shlomo}. We further demonstrated that although better annealing of glasses can lead to a significant relieving of internal stresses, its effect on the form of the density of vibrational modes is qualitatively different compared to artificially reducing the internal stresses. This difference highlights once again the marginal nature of structural glasses and the importance of the self-organizational processes taking place during cooling from the liquid state. An interesting question to be addressed next is whether the gapless $\omega^4$ law observed under mild annealing is robust to extremely slow annealing, e.g.~in vapor deposited glasses~\cite{vapor_deposited_glasses_prl_2017}, or glasses created by the swap monte-carlo method~\cite{swap_prx}.

\textit{Acknowledgments.--} E.~L.~acknowledges support from the Netherlands Organisation for Scientific Research (NWO) (Vidi grant no.~680-47-554/3259). E.~B.~acknowledges support from the Minerva Foundation with funding from the Federal German Ministry for Education and Research, the William Z.~and Eda Bess Novick Young Scientist Fund and the Harold Perlman Family.


%

\end{document}